\documentclass[runningheads]{llncs}
\usepackage[T1]{fontenc}
\usepackage{graphicx}
\usepackage{orcidlink}
\usepackage{subcaption}

\begin{document}
\title{Energy Consumption of TLS, Searchable Encryption and Fully Homomorphic Encryption}
\titlerunning{Energy Consumption of TLS, Searchable Encryption, and FHE}
%
\author{Marc Damie\inst{1,2}\orcidlink{0000-0002-9484-4460}\thanks{Corresponding author: \email{m.f.d.damie@utwente.nl}} \and
    Mihai Pop\inst{1} \and
    Merijn Posthuma\inst{1}}

\authorrunning{M. Damie et al.}
%
\institute{University of Twente, The Netherlands \and
    Inria, France}
\maketitle              
\begin{abstract}
    Privacy-enhancing technologies (PETs) have attracted significant attention in response to privacy regulations, driving the development of applications that prioritize user data protection.
    At the same time, the information and communication technology (ICT) sector faces growing pressure to reduce its environmental footprint, particularly its energy consumption.
    While numerous studies have assessed the energy consumption of ICT applications, the environmental impact of cryptographic PETs remains largely unexplored.

    This work investigates this question by measuring the energy consumption increase induced by three PETs compared to their non-private equivalents: TLS, Searchable Encryption, and Fully Homomorphic Encryption (FHE).
    These technologies were chosen for two reasons.
    First, they cover different maturity levels --from the widely deployed TLS protocol to the emerging FHE schemes-- allowing us to examine the influence of maturity on energy consumption.
    Second, they each have well-established applications in industry: web browsing, encrypted databases, and privacy-preserving machine learning.

    Our results reveal highly variable energy consumption increases, ranging from $2\times$ for TLS to $10\times$ for Searchable Encryption and $100{,}000\times$ for FHE.
    Our experiments demonstrate a simple and reproducible methodology, based on existing open-source software, to quantify the energy consumption of PETs.
    They also highlight the wide spectrum of energy demands across technologies, underscoring the importance of further research on sustainable PET design.
    Finally, we discuss orthogonal research directions, such as hardware acceleration, to outline promising directions toward sustainable PETs.

\end{abstract}
%
%
%
\section{Introduction}
Over the last ten years, awareness about privacy issues has significantly increased.
Legislations such as the European GDPR (General Data Protection Regulation) marked a turning point by incentivizing practitioners to develop applications considered as ``private by design.''
Pushed by these legal incentives, the research community has proposed many novel privacy-enhancing technologies (PETs); enabling to implement existing applications without requiring the user to reveal their private data.

However, information and communication technologies (ICTs) are also coping with another major challenge: reducing their environmental footprint; in particular their carbon emissions.
In their fight against climate change, many governments are willing to reduce their carbon footprint, and all industries (including ICTs) attempt to contribute to the carbon emission reductions.

Several existing works started exploring this problem either by estimating the carbon footprint of the ICT industry \cite{gupta_chasing_2022} or by measuring the footprint of specific applications.
Measurements are essential to identify the most energy-efficient technologies in order to fulfil carbon emission reduction goals.

\paragraph{Related works}
Existing works have measured the environmental footprint of various ICTs: online advertising \cite{albasir_experimental_2014}, network communications \cite{ficher_assessing_2021}, video streaming \cite{afzal_survey_2024}, video calls \cite{berthoud_evaluation_2022}, Machine Learning (ML) \cite{georgiou_green_2022,henderson_towards_2020,schmidt_codecarbon_2021}, and blockchain \cite{bada_towards_2021,sedlmeir_energy_2020}.
However, there has been limited research on the impact of PETs.

On the one hand, several works \cite{miranda_tls_2011,naylor_cost_2014} have studied the energy consumption of the TLS protocol used to secure various web protocols (including HTTP).
This represents important related works as TLS is a form of ``soft'' PET (i.e., a PET providing data security and processing data with consent \cite{deng_privacy_2011}).

On the other hand, various ML paradigms have been studied under the lens of energy consumption, including differentially private ML \cite{naidu_towards_2021,de_reus_energy_2023} and Federated ML \cite{guerra_cost_2023,savazzi_energy_2022}; two paradigms related to privacy-preserving ML.
Our work is complementary to these research works because we notably measure the consumption of cryptographic techniques applied to privacy-preserving ML.

\paragraph{Gap in the literature}
Despite these preliminary works, the debate about environmental sustainability is absent from the cryptography community.
This absence is particularly problematic because cryptographic PETs induce computation and communication overheads compared to their ``non-private'' equivalents.
While privacy could be seen as a value to protect at all costs, the emission reduction goals may require to design PETs offering the best trade-off between privacy and carbon emissions.

Unfortunately, the literature provides no evaluation of the environmental footprint of cryptographic PETs.
Such measurements have become essential to find the best trade-off between privacy and carbon emissions.

\paragraph{Our contributions}
We present a simple and reproducible methodology (used notably in ML) to measure the energy consumption of PETs.
Using this methodology, we measure the energy consumption of three PETs: TLS, Searchable Encryption, and Fully Homomorphic Encryption.
These three PETs were selected because they combine practical relevance with different maturity levels.
On the one hand, they have well-identified industrial applications: web browsing, encrypted databases, and privacy-preserving ML.
On the other hand, they have different maturity levels from the widely deployed TLS protocol to the emerging FHE schemes.
Our experiments highlight the relative increase between the PET and its non-private equivalent.
We observe increases ranging from a twofold overhead with TLS to a 100{,}000-fold overhead with FHE.
Finally, we discuss several orthogonal questions (including hardware acceleration) to identify promising directions for sustainable PET design.

Our goal is not to categorize PETs between environmentally acceptable and unacceptable.
To fulfil carbon emission goals, it is not mandatory to reduce the carbon emissions of all technologies: some highly energy-consuming technologies can be considered ``worth the emissions'' because their service is essential.
Our work simply provides orders of magnitude enabling decision makers to assess the privacy-efficiency trade-offs of specific services.
Such figures have become essential in the public debate as press articles rely on estimations and measurements from the scientific literature to provide high-level perspectives on sustainability issues \cite{griffiths_why_2020}.

\section{Methodology}
\label{sec:methodology}

Our work adopts a methodology commonly used in scientific studies (notably in ML research \cite{de_reus_energy_2023,schmidt_codecarbon_2021}) to measure the energy consumption of various algorithms.  
This section presents the methodology, its underlying assumptions, and the general philosophy guiding its use in our study.

\subsection{Life-Cycle Assessment}
Life-Cycle Assessment (LCA) is a standard methodology \cite{hauschild_life_2017} to measure the environmental footprint associated with all stages of a product's life.
Such stages can include anything from the manufacturing to the recycling.
This approach enables to estimate the energy consumption, carbon emissions, and resource consumption in a comprehensive manner.

An LCA requires \textbf{defining a scope} based on an analysis goal; this will specify what is included in and excluded from the analysis.
Let us assume that we want to compare the footprint of two objects.
If the two objects use the same materials and end-of-life processes, the material extraction and the end of life are not necessary in the analysis.
Indeed, these stages affect equally the two objects, so they will not change the comparison (i.e., our objective).


The analysis goal also defines an analysis metric which quantifies the impact of a product.
The two most common metrics are the energy consumption (in kWh) and the carbon emissions (in kg eq. CO\textsubscript{2}) \cite{schmidt_codecarbon_2021}.
However, some analyses also quantify other variables such as water consumption \cite{berger_water_2010}.

\paragraph{LCA and digital services}
While LCA initially focused on physical objects, it is possible to extend this approach to digital services.
However, LCA on such services focuses mostly on the service usage.
For example, for an ML service, an LCA studies the impacts of the model training \cite{henderson_towards_2020,luccioni_estimating_2023,schmidt_codecarbon_2021} and inference \cite{faiz_llmcarbon_2024}.

One may wonder why the hardware is not always included in the setup.
Even though hardware is a significant part of the ICT footprint \cite{gupta_chasing_2022}, the hardware-related costs are irrelevant in the scope of these works.
These works \cite{henderson_towards_2020,schmidt_codecarbon_2021} (like ours) want to identify the most energy-efficient services and execute them on a standard hardware.
In other words, the hardware-related footprint is excluded because they want to minimize the consumption of the service usage.

Like most related works on the environmental footprint of ICTs \cite{georgiou_green_2022,henderson_towards_2020,miranda_tls_2011,naidu_towards_2021} \cite{naylor_cost_2014,de_reus_energy_2023,savazzi_energy_2022,schmidt_codecarbon_2021}, our LCA focuses on the service usage (especially the energy consumption).
Therefore, our experiments quantify the environmental impact in terms of \textbf{energy consumption}.

\subsection{Software-based measurement}
\label{subsec:soft-based-framework}
Our scope requires measuring the energy consumption induced by a software.
A natural approach is to use a power meter.
While reliable, such hardware-based approaches are inconvenient as they limit the reproduction.

On the contrary, recent works \cite{jay_experimental_2023,khan_rapl_2018,schmidt_codecarbon_2021} have promoted software-based approaches to measure the energy consumption.
Software-based measurements are valuable as they simplify the reproduction of an experiment: they simply require a package installation.
In particular, Khan et al. \cite{khan_rapl_2018} showed that Intel's Running Average Power Limit (RAPL) was a powerful tool to measure the power consumption of a machine.
RAPL-based measurement consists in polling the RAPL interface every $x$ seconds and extrapolate the consumption based on these discrete measurements.
Such measurements were notably used in ML research \cite{schmidt_codecarbon_2021}.
Recently, Jay et al. \cite{jay_experimental_2023} highlighted that RAPL induces a low overhead.

Considering the advantages of RAPL, like related ML works \cite{schmidt_codecarbon_2021}, we rely on this software-based approach to measure energy consumption.
Schmidt et al. \cite{schmidt_codecarbon_2021} extended this approach to include GPU and RAM consumption.

From the energy consumption, note that it is possible to estimate the carbon emission.
To compute the carbon emissions, one needs to multiply the energy consumption to the ``carbon intensity'' of the country in which the server/client is located.
The carbon intensity corresponds to the amount of greenhouse gas emitted per kWh of electricity produced.
This information is directly available in some public databases (such as those of the Ember and Energy Institute).
In other words, the energy consumption and the energy emissions are perfectly correlated, so we do not provide it.

\paragraph{Indirect communication-related energy consumption}
Our experiments exclude the energy consumption indirectly induced by network communications.
While it is straightforward to measure the energy consumption of a server or client, estimating the total energy required to transmit information between two devices is more challenging.
To the best of our knowledge, no study has yet estimated the complete footprint of client-server communications, as this would require computing the amortized energy consumption of all network devices involved (e.g., routers, switches, antennas, wires) for transmitting a bit of information.
Ficher et al.~\cite{ficher_assessing_2021} made an initial contribution by studying communication between two servers within the same network backbone, but their work does not provide a generalizable result.

Anyway, the technologies studied in our work induce only minor communication overheads.
For example, in HTTPS web browsing, the size of the ciphertext sent to the server is comparable to the plaintext, apart from a one-time secret-key exchange.
Thus, their communication overhead is negligible compared to that of ``communication-intensive PETs,'' such as multiparty computations.

\subsection{Is the runtime a good proxy for the energy consumption?}
As research papers typically benchmark the runtime of their algorithms, one may want to extrapolate the energy consumption based on such measurements.
While runtime is correlated to energy consumption, other parameters influence it.
Thus, we cannot trivially deduce the energy consumption based on the runtime.

To reduce their runtime, research works (including in PETs \cite{klemsa_parmesan_2023,mazzone_efficient_2024}) parallelize their computation across multiple cores.
When parallelized, the energy consumption of an algorithm is proportional to the number of cores in use.
Thus, a runtime can be divided by 4, 16, or even 32 via parallelization, but the energy consumption would remain similar because spread over several cores.
Similarly, powerful CPUs and GPUs can also reduce the runtime, but induce a higher energy consumption than smaller chips.

This observation shows that specialized energy measurement techniques (such as RAPL \cite{khan_rapl_2018}) are essential to quantify the consumption and cannot be obtained using simpler methods such as runtime measurement.
For example, RAPL-based measurements account for the energy consumption of all cores.

In conclusion, we provide a complementary perspective to existing PET papers because we report energy measurements that cannot be \emph{precisely} extrapolated from existing experimental results.
Our work echoes with recent ML work \cite{henderson_towards_2020} that promoted a systematic reporting of carbon footprint and energy consumption in ML papers (in addition to existing runtime measurements).

\section{Results}
\label{sec:experiments}

This section measures the \textbf{relative difference} between the energy consumption of three PETs and their non-private equivalents.  
Since it is not feasible to benchmark all PETs in a single study, we focus on three representative technologies selected for their maturity and industrial relevance: FHE, Searchable Encryption, and TLS.  
On the one hand, these technologies cover different maturity levels: TLS is widely deployed, Searchable Encryption has been actively studied for two decades, and FHE has grown rapidly since the introduction of more efficient schemes \cite{cheon_homomorphic_2017,chillotti_tfhe_2019} less than ten years ago.  
On the other hand, they each have well-identified applications: encrypted machine learning, encrypted databases, and secure web browsing.  
With this scope, our goal is both to initiate research on the environmental footprint of PETs and to demonstrate an accessible and reproducible methodology.  

This scope naturally leaves room for future studies on other PETs, such as multiparty computation (MPC), functional encryption, or trusted execution environments (TEE).
These technologies raise distinct challenges --ranging from massive communication overheads in MPC to indirect hardware-related costs in TEE (see Section \ref{subsec:crypto-hardware})-- that call for dedicated investigation.

\subsection{Experimental setup}
We execute our experiments on a dedicated server with 32 GB, an Intel Xeon-E3 1245 v5, and no GPU.
We run the client and the server on the same machine.

As the hardware can slightly change the energy consumption of an algorithm, one may wonder whether we should execute our experiment on different devices.
However, remember that we mainly want to measure the relative consumption difference between private and non-private deployments.
A standard hardware change would influence the absolute values, but the relative difference should not be significantly changed.
This measurement on a single hardware platform aligns with common practices in similar studies on ML algorithms \cite{de_reus_energy_2023,guerra_cost_2023}.

The only hardware that could significantly influence a PET benchmark is cryptographic accelerators.
We leave them out-of-scope because a dedicated analysis is necessary to estimate the production cost of this hardware \emph{only needed} in the private deployment.
Section \ref{subsec:crypto-hardware} further discusses the use of such hardware.

We use the software CodeCarbon \cite{schmidt_codecarbon_2021} to measure the energy consumption.
This software uses a RAPL-based measurement, and measures RAM, CPU, and GPU consumption.
We configured CodeCarbon to poll the energy consumption every 1 ms (like in \cite{jay_experimental_2023}).
Our measurement aggregates the computational costs of the client and the server.

Our experiments systematically use the default configuration proposed in the documentation of the tested software.
Thus, the measurements reported in our paper correspond to the expected default behavior.
For implementation details, we refer to our \textbf{publicly available codebase}:

\url{https://github.com/MarcT0K/privacy-carbon-experiments}

\subsection{Fully Homomorphic Encryption (FHE)}
FHE is a type of cryptographic scheme \cite{cheon_homomorphic_2017,chillotti_tfhe_2019} allowing to perform arithmetic operations on encrypted data.
With the increasing use of ML, there is a growing concern regarding the use of personal data.
To address this problem, FHE has appeared as a nice solution to execute traditional ML pipelines under encryption.

\emph{Software libraries.}
Our goal is to measure the energy consumption of an encrypted ML operations compared to the equivalent plaintext operations.
We use the software library Concrete ML developed by the company Zama.
This library is stable and promoted by the company for real-world use cases.
We compare the encrypted operations provided by Concrete ML to the plaintext ML performed by Scikit Learn (a popular ML framework).
Note that Scikit Learn is also used as baseline by Zama in their accuracy benchmarks.

Note that Concrete ML only supports TFHE  \cite{chillotti_tfhe_2019}, but there exists another popular FHE scheme in the literature, namely CKKS \cite{cheon_homomorphic_2017}.
Unfortunately, existing CKKS implementations are much more basic than Concrete ML, and do not support high-level operations like ML.
While Concrete ML is designed for industrial applications and maintained by a company, other FHE implementations are not as mature.
Thus, we only experiment on TFHE, because other implementations either provide insufficient functionalities, or are too immature (i.e., providing an unfair comparison to an industrial software like Concrete ML).

Anyway, our work does not aim to identify the best FHE implementation, or even the best privacy-preserving ML framework in general.
We want to estimate an order of magnitude of the energy consumption increase induced by an FHE-based application compared to its plaintext equivalent.
Moreover, the FHE literature is particularly active, so reproducibility studies would be needed in the upcoming years to measure the evolution.

\emph{Data.}
We use Scikit Learn to generate synthetic datasets with varying number of features.
In ML, a sample is a data point included in a dataset, and the features are the dimensions/characteristics of a data point.

\subsubsection{Encrypted Machine Learning inference}
Among all ML operations, ML inference (i.e., evaluating data on a trained ML model) is a particularly important operation for practitioners.
Indeed, many companies have trained a powerful model (e.g., a Large Language Model) and provide ``Inference as a Service'': customers send input data and the company returns the model output.
To enhance privacy, researchers have designed encrypted ML frameworks \cite{chillotti_tfhe_2019}.

\begin{figure}[t]
    \centering
    \includegraphics[width=0.7\linewidth]{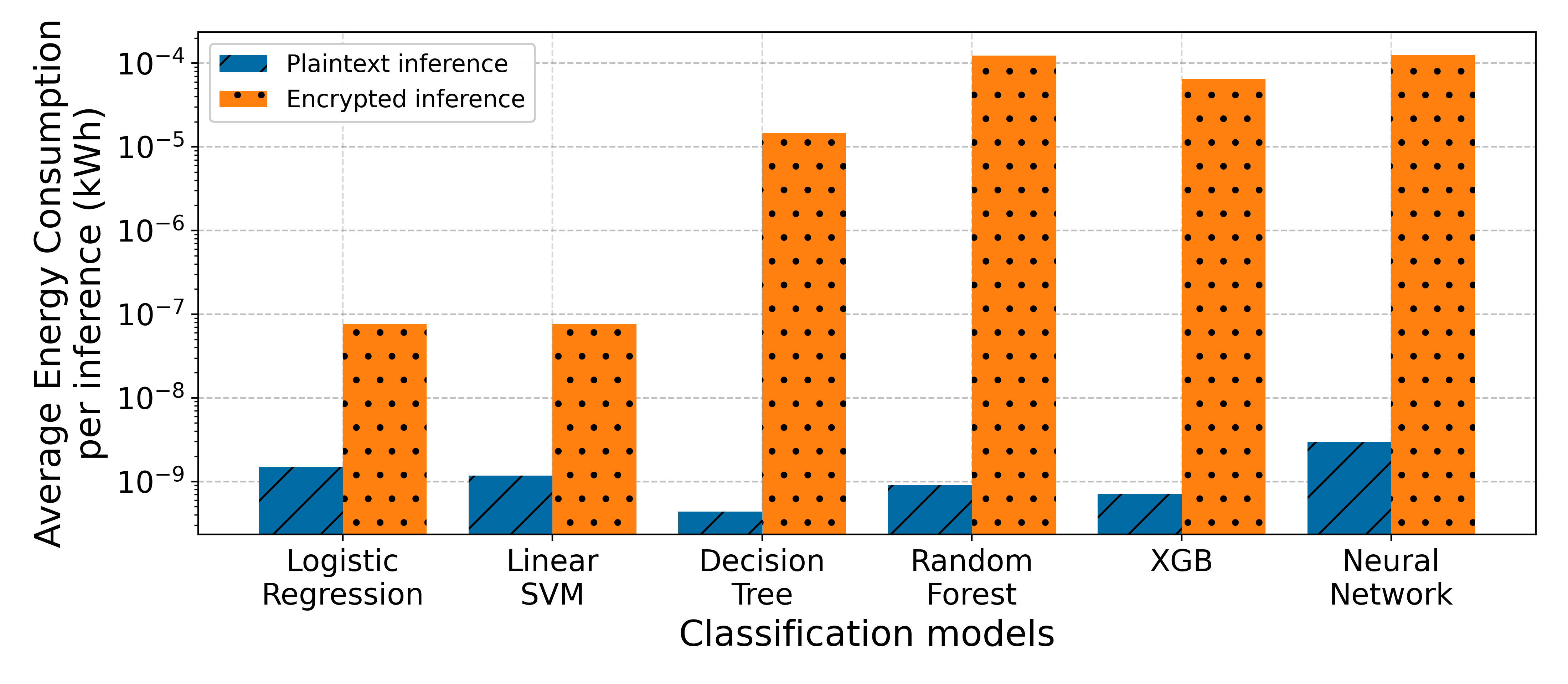}
    \includegraphics[width=.7\linewidth]{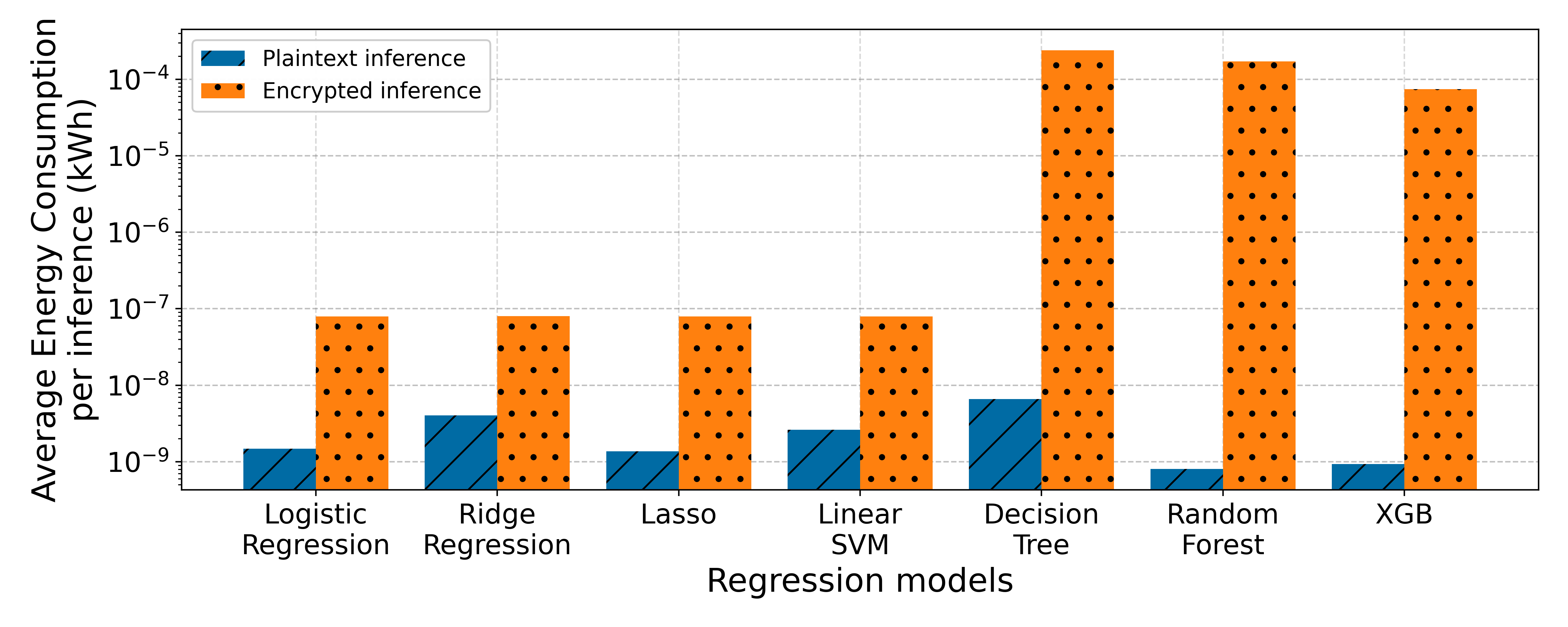}
    \caption{Average energy consumption of encrypted and plaintext ML inference using various classification models (100 samples, 30 features).}
    \label{fig:ml-inference}
\end{figure}

\emph{Results.}
Figure \ref{fig:ml-inference} presents measurements for various classification and regression models.
These results are average measurements over 100 samples; each sample having 30 features.
On linear models (e.g., Logistic regression), the encrypted inference is $100\times$ more expensive than the plaintext inference.
On tree models (Decision Tree, Random Forest, and XGB), the encrypted inference is at least $100,000\times$ more expensive than the plaintext inference.
Encrypted inference on neural networks is also $100,000\times$ more expensive than plaintext.

\begin{figure}[t]
    \centering
    \begin{subfigure}{0.49\linewidth}
        \includegraphics[width=\linewidth]{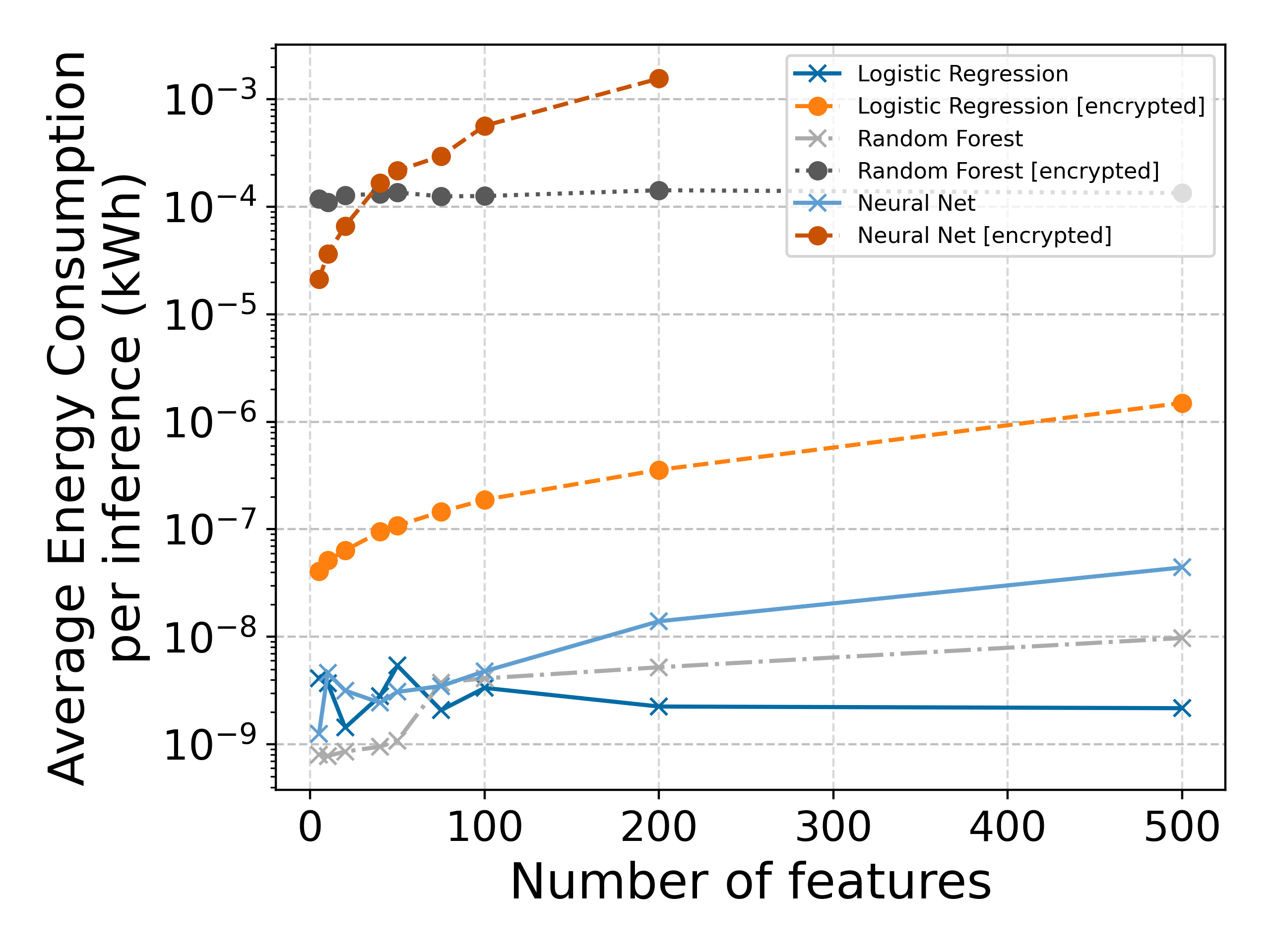}
        \caption{Varying nb. of features / 100 samples}
        \label{fig:ml-inference-varying-features}
    \end{subfigure}
    \begin{subfigure}{0.49\linewidth}
        \includegraphics[width=\linewidth]{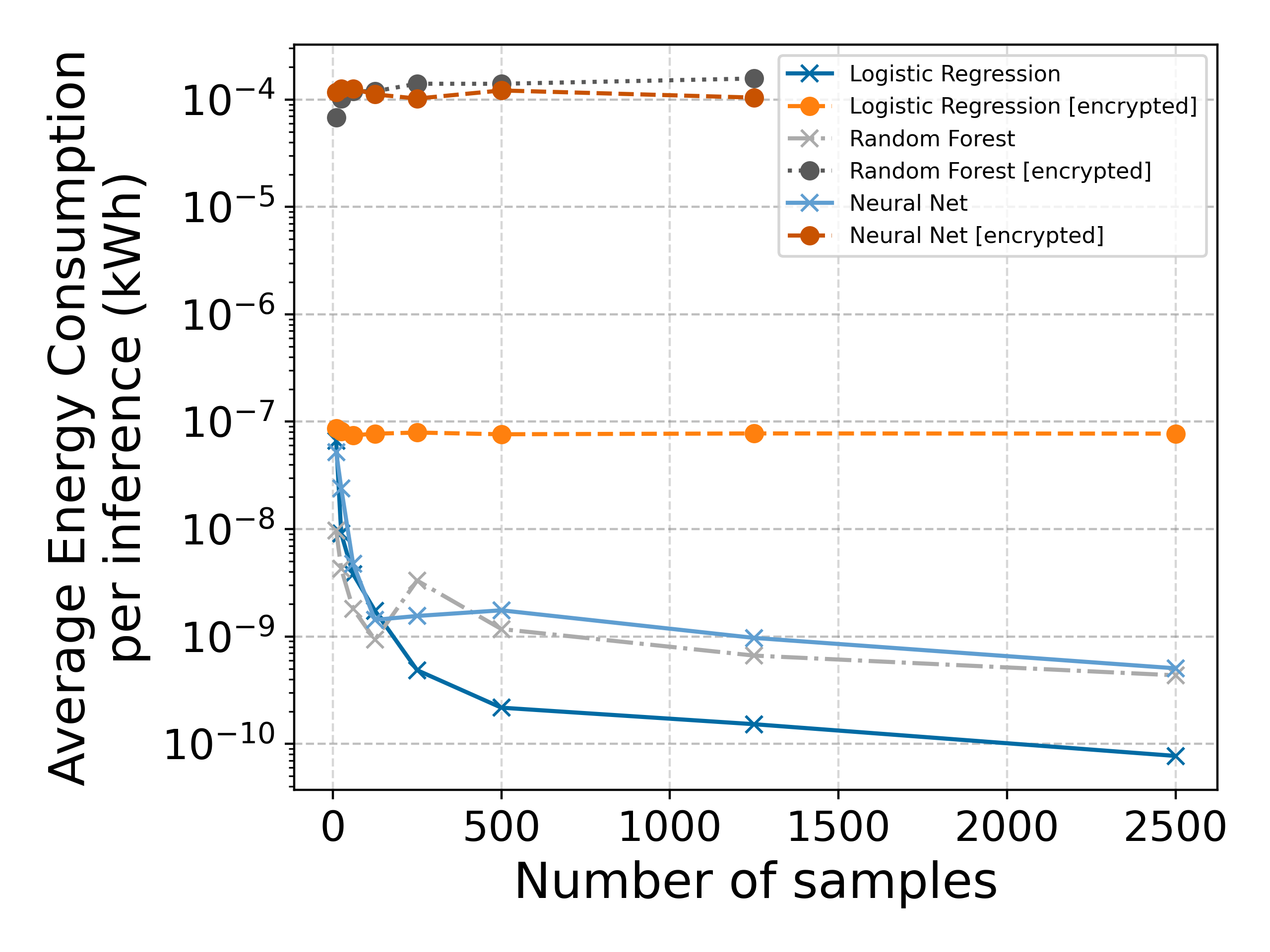}
        \caption{Varying nb. of samples / 30 features}
        \label{fig:ml-inference-varying-samples}
    \end{subfigure}
    \caption{Average energy consumption of encrypted and plaintext ML inference for varying number of features and samples.}
\end{figure}

To better understand this overhead, Figure \ref{fig:ml-inference-varying-features}  studies the influence of the number of features on the consumption of three popular classification models: logistic regression, random forest, and neural networks.
We observe that the scaling of the encrypted inferences does not match perfectly the scaling of the plaintext inferences.
First, the encrypted logistic regression consumption increases relatively faster than the cost of the plaintext inference.
Second, the encrypted neural network seems to have the same scaling pattern as the encrypted logistic regression.
The encrypted neural network does not have results for more than 200 features because we stopped inferences taking more than several hours (to limit the energy consumption of our experiments).
Third, the encrypted random forest consumption is approximately constant.
This behavior seems to be an artifact coming from the fact that Concrete ML transforms decision trees into matrix multiplications (inducing significant constant costs).

Figure \ref{fig:ml-inference-varying-samples} illustrates the consumption in function of the number of samples.
On the one hand, we see that plaintext algorithms are optimized to process batches of data points: the average cost is amortized when the number of samples grows.
On the other hand, we observe on the logistic regression that the encrypted inference does not provide a similar amortization.
Like on Figure \ref{fig:ml-inference-varying-features}, we stopped experiments requiring several hours (which explains the smaller number of results for the random forest and neural network).

\emph{Discussions.}
Our experiments show that encrypted ML inference induces significant overheads.
However, this PET is still relatively recent, so there is room for optimization.
For example, Ko et al. \cite{ko_silentwood_2024} recently presented an encrypted XGB $100\times$ more efficient than Concrete ML's XGB.
This improvement resonates with Figure \ref{fig:ml-inference-varying-features} that identified a major constant factor for encrypted Random Forest (i.e., a tree-based model related to XGB).
Our benchmark could be re-executed once Concrete ML integrates improved algorithms.

\subsubsection{Encrypted Machine Learning training}
Next to inference, ML training is the other operation attracting a lot of privacy concerns.

\begin{figure}[t]
    \centering
    \begin{subfigure}{0.49\linewidth}
        \includegraphics[width=\linewidth]{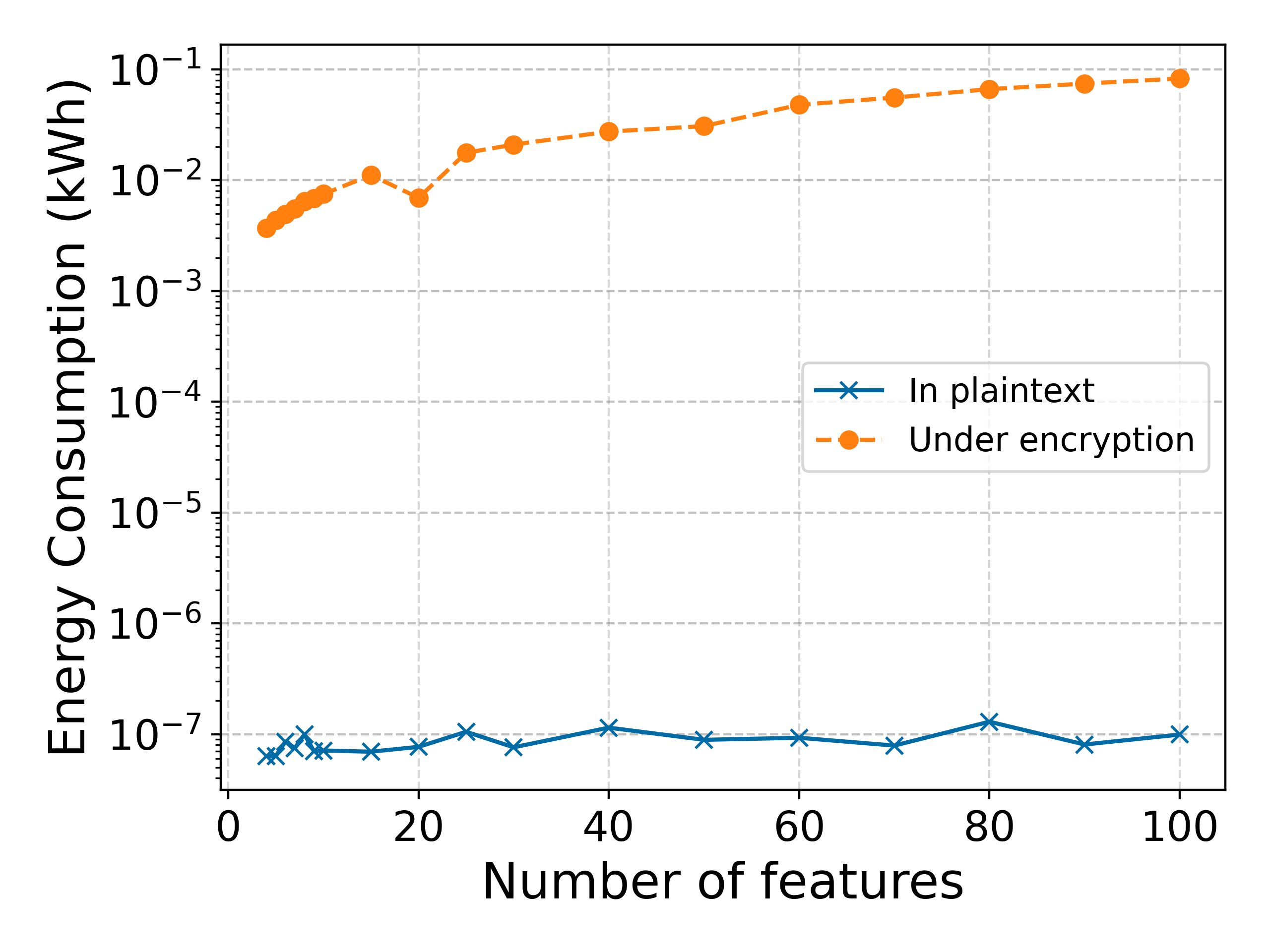}
        \caption{Varying nb. of features / 300 samples}
        \label{fig:ml-training-varying-features}
    \end{subfigure}
    \begin{subfigure}{0.49\linewidth}
        \includegraphics[width=\linewidth]{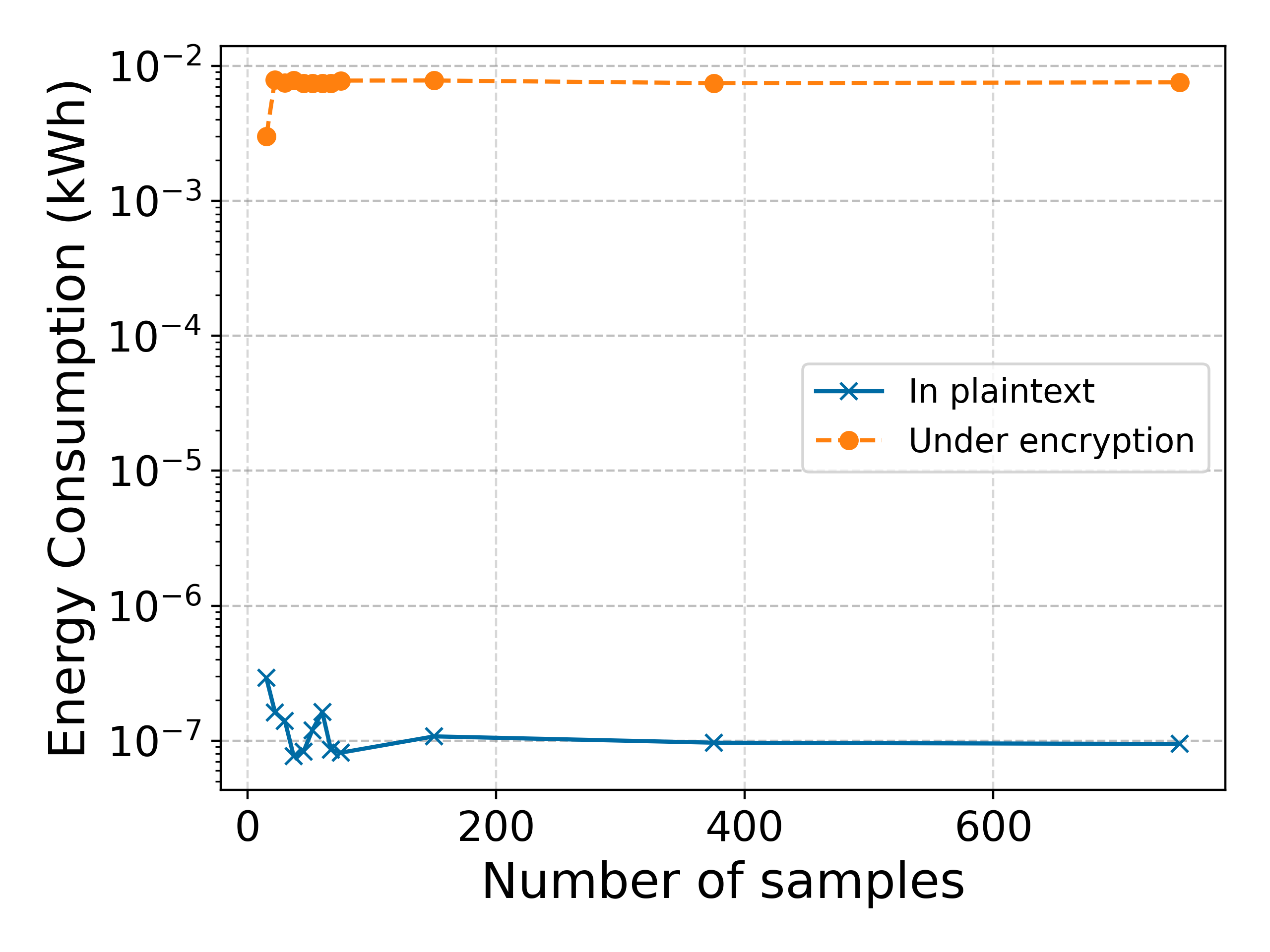}
        \caption{Varying nb. of samples / 10 features}
        \label{fig:ml-training-varying-samples}
    \end{subfigure}
    \caption{Energy consumption of encrypted and plaintext ML training of a logistic regression for varying numbers of features and samples}
\end{figure}

\emph{Results.}
Figure \ref{fig:ml-training-varying-features} compares the encrypted and plaintext training (of a logistic regression) for a varying number of features.
The encrypted training is at least $100,000\times$ more expensive and its energy consumption increases faster with the number of features.

Contrary to inference, the logistic regression training is not influenced by the number of samples.
Its training algorithm is iterative and process a fixed-sized batch during each iteration.
Thus, its cost only depends on the batch size and number of iterations.
Figure \ref{fig:ml-training-varying-samples} confirms this behavior experimentally.

\emph{Discussions.}
Like encrypted ML inference, secure training induces a massive consumption increase.
Similarly, there is room for optimizations.

However, other (non-cryptographic) approaches to privacy-preserving ML training exist.
For example, de Reus et al. \cite{de_reus_energy_2023} studied the energy consumption of synthetic data generation in ML context.
Synthetic data generation is a pre-processing step used in privacy-preserving statistics and ML: instead of encrypting its private data, the data owner generates synthetic data based on its data and can transmit the synthetic data in plaintext.
In this paradigm, the server can then process plaintext data without privacy issues as it has only access to synthetic data (and not the initial private data).
Some techniques such as \cite{jordon_pate-gan_2018} ensure that the synthetic data does not leak private information.

De Reus et al. \cite{de_reus_energy_2023} showed that their studied synthetic data generation requires 1 Wh to generate a synthetic dataset based on the adult dataset (14 features), and 0.01 Wh to train the logistic regression.
They also used a RAPL-based energy measurement, so our results are comparable.
In comparison, Figure \ref{fig:ml-training-varying-features} shows that the encrypted ML training requires $10\times$ more energy for a training on a similar dataset.

This comparison emphasizes that the PET choice has a major impact on the energy consumption.
It also highlights an interesting future work: comparing all privacy-preserving ML paradigms to identify the most energy-efficient one.

\subsection{Searchable Encryption}
With the rise of cloud services, vast amounts of personal data are stored on outsourced databases, raising privacy concerns since providers may not be fully trusted.
To solve this issue, researchers introduced searchable encryption \cite{curtmola_searchable_2006}, a database model that enables encrypted data storage and querying while preventing the provider from accessing data or query content.

\emph{Software libraries.}
This experiment\footnote{Due to hardware issues (unrelated to the experiment), we executed this experiment on a MacOS machine while the other experiments are executed on a Debian server.} uses SWiSSSE \cite{gui_swissse_2024} as baseline for encrypted database.
SWiSSSE is an encrypted database system comparable to Redis (a traditional database specialized in key-value storage).

Gui et al. \cite{gui_swissse_2024} already compared the runtime of SWiSSSE to Redis.
We reproduce their benchmark using new metrics: energy.

\emph{Data.} Like in \cite{gui_swissse_2024}, we populate the database using the Enron email dataset.

\begin{figure}[t]
    \centering
    \includegraphics[width=.6\linewidth]{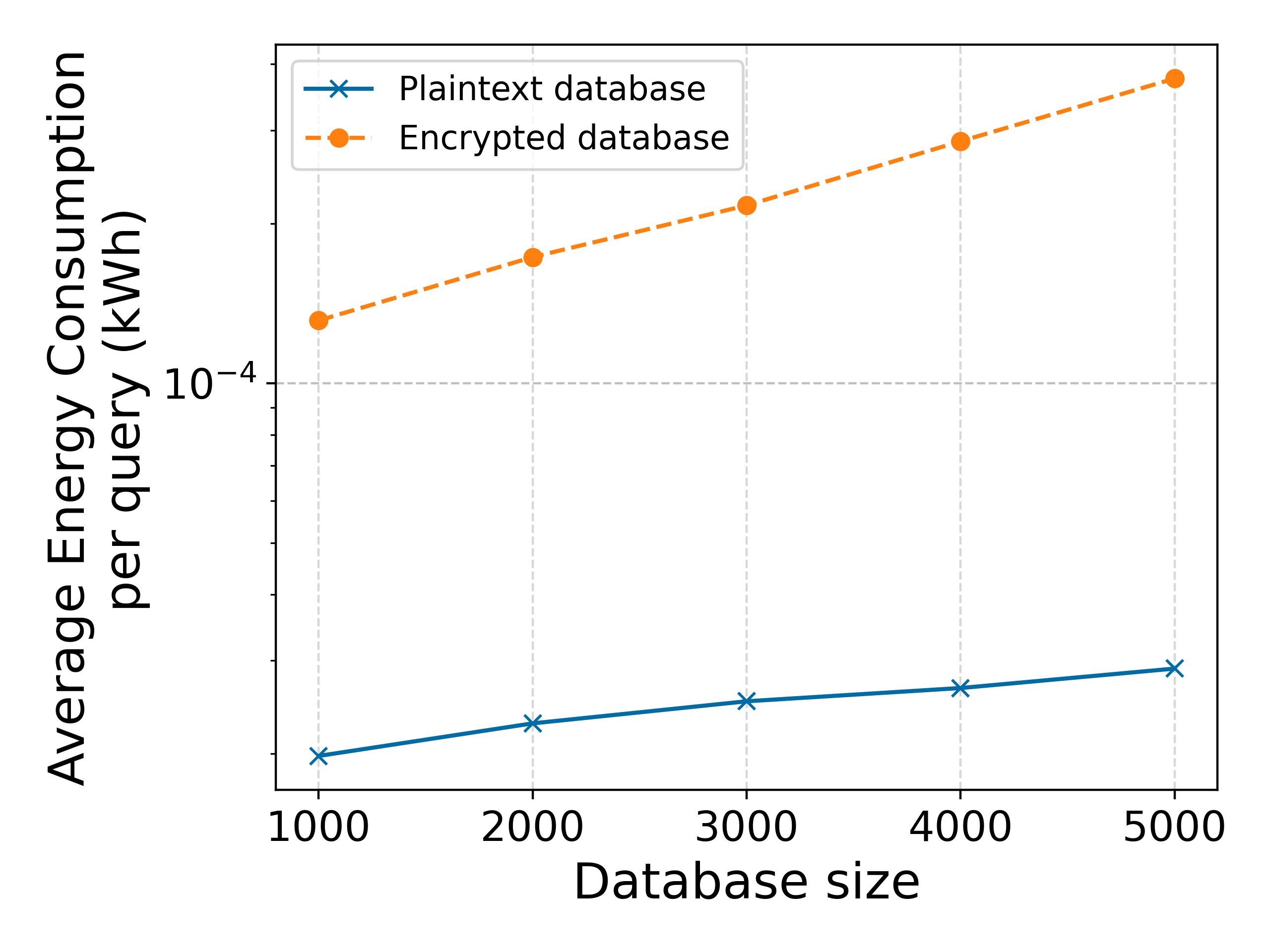}
    \caption{Average energy consumption of encrypted and plaintext queries (SWiSSSE vs. Redis) for varying database sizes (1000 queries).}
    \label{fig:sec-db-varying-db-size}
\end{figure}

\emph{Results.}
Figure \ref{fig:sec-db-varying-db-size} shows the results of this benchmark for varying database sizes.
We observe that the consumption of the encrypted database is nearly $10\times$ higher than the consumption of the plaintext database.
Moreover, the consumption of the encrypted database increases slightly faster in function of the database size.

\emph{Discussions.}
Our benchmark relies on SWiSSSE, which is a research prototype.
Ideally, we would like to reproduce the results on professional products such as the ``Queryable Encryption'' plugin available in MongoDB.
Unfortunately, part of this plugin requires a premium license; limiting its access and hindering reproducibility.
We demonstrated our methodology on SWiSSSE, and leave for future work the evaluation of MongoDB Queryable Encryption.

\subsection{TLS}
\label{subsec:https}
HTTP (HyperText Transfer Protocol) is the foundational protocol used for data exchanges on the web.
It operates as a plaintext protocol; making the transmitted data susceptible to eavesdropping and tampering.
In contrast, HTTPS integrates HTTP with the Transport Layer Security (TLS), encrypting data to ensure confidentiality and integrity.
HTTPS protects sensitive data, such as login credentials and payment information, from being intercepted or altered.

\emph{Software libraries.}
We use NGinx as web server with TLS 1.3 and implemented a basic Python web client using the \texttt{requests} library.
We compute the average energy consumption and carbon emissions over 1000 Web requests.

\emph{Data.}
We run this experiment on five different websites: Wikipedia (Simplified English), New York Times, MDN, Mastodon Technical Blog, and xkcd.
These websites are quite diverse: Wikipedia is an encyclopedia, New York Times a media, MDN a web developer documentation, Mastodon Blog is a blog, and xkcd is a minimal website publishing amusing comic strips.
Wikipedia source files are publicly available, so we downloaded the ``simplified English'' archive from April 2007.
We use the tool HTTrack to automatically download the static files (i.e., HTML, CSS, JS, and images) of the other websites.

\begin{figure}[t]
    \centering
    \includegraphics[width=.6\linewidth]{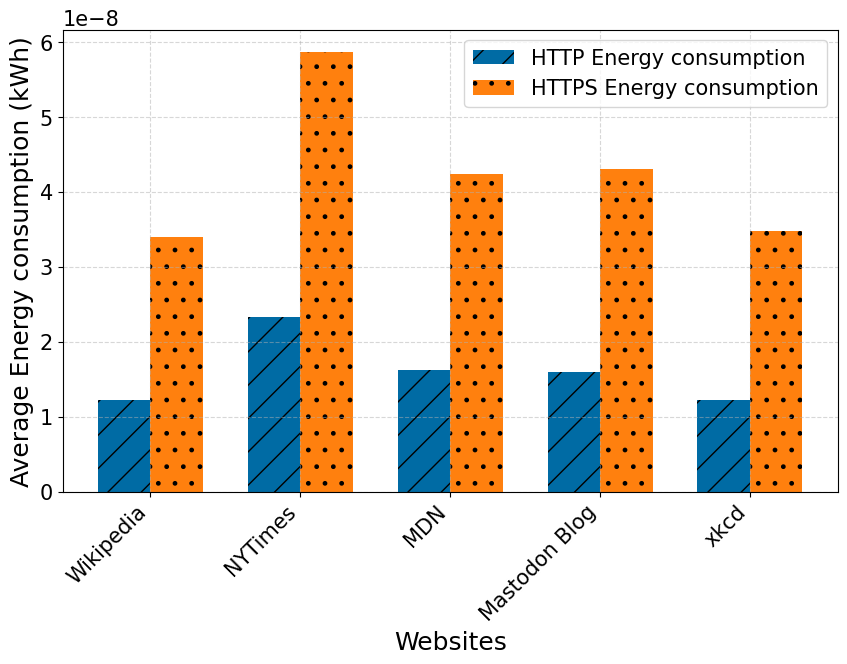}
    \caption{Average energy consumption of HTTP and HTTPS requests on five different websites (1000 requests).}
    \label{fig:https}
\end{figure}

\emph{Results.}
Figure \ref{fig:https} represents the average energy consumption over 1000 requests.
The relative increase ranges from 152\% for NYTimes to 182\% for xkcd.

Unfortunately, we cannot compare our results to existing works on the energy consumption of TLS \cite{miranda_tls_2011,naylor_cost_2014}.
While we analyze the energy consumption of the whole HTTPS protocol, these works had a completely different goal: estimating the impact of HTTPS on a smartphone battery life.
Thus, these works have only measured client-side energy consumption.

Moreover, Miranda et al. \cite{miranda_tls_2011} focused on the energy consumption of TLS only; they do not provide a relative difference between HTTP and HTTPS.
Finally, Naylor et al. \cite{naylor_cost_2014} focused on specific Web applications (e.g., Youtube); contrary to our experiments that considered an easily reproducible setup: Python web client and static HTML files.
Naylor et al. \cite{naylor_cost_2014} also integrated some network-related costs (e.g., WiFi communications).
These major divergences between our experimental setup and \cite{naylor_cost_2014} prevent any naive comparison.

\emph{Discussions.}
Our experiments only simulate static websites, and exclude any possible back-end operations (e.g., database interactions), even though such operations are common in Web applications.
To assess the footprint of a Web application, an experiment would include both the cost related to HTTP(S) and the back-end costs.
However, this is not our goal: we only want to compare HTTP to HTTPS.
Hence, we exclude any backend work to focus solely on the consumption of HTTP data transfer.

\section{Orthogonal discussions}
\label{sec:discussions}

\subsection{The hidden footprint of cryptographic hardware}
\label{subsec:crypto-hardware}
Energy consumption can be significantly reduced by using cryptographic accelerators \cite{bisheh-niasar_cryptographic_2021}.
These chips can perform cryptographic operations much more efficiently than standard CPUs and GPUs.

However, the LCA must include the manufacturing impact, as this specialized hardware requires the production of additional hardware (in addition to the standard CPU).
Since manufacturing accounts for most of the ICT sector's carbon footprint \cite{gupta_chasing_2022}, using cryptographic accelerators does not necessarily reduce the overall carbon footprint.

Trusted Execution Environments (TEEs) \cite{sabt_trusted_2015} are another example of cryptographic hardware with an indirect environmental footprint that should not be ignored.
TEEs are now widely used in efficient secure protocols because they allow operations on confidential data without relying on expensive primitives like homomorphic encryption.
For example, Signal uses TEEs in its Contact Discovery protocol.

As their use grows, more manufacturers now include TEEs by default in their products.
However, the added manufacturing cost still represents an overhead that needs to be estimated.
Like cryptographic accelerators, estimating this cost is difficult, as the necessary data is rarely made public by production facilities.
This gap then represents a promising and interesting future work.

\subsection{Privacy-Energy-Functionality trilemma}
Our work initially examined the relationship between privacy and energy emissions in an isolated manner.
However, our experiments indirectly revealed a broader trilemma involving privacy, energy emissions, and functionality.

The energy consumption of cryptographic PETs is tied to their functionality; more complex functionalities generally require more computationally expensive cryptographic primitives, leading to higher emissions.
This phenomenon is evident in Figure \ref{fig:ml-inference}.
While encryption always increases the consumption, the extent of this increase varies significantly across ML models.
For instance, logistic regression exhibits a 100-fold increase, whereas encrypted neural networks result in a 100,000-fold increase.
Consequently, simpler models like logistic regression lead to a smaller consumption.

Reducing functionality can therefore serve as an effective strategy to mitigate the overhead induced by the privacy enhancement.

\subsection{Mitigating overhead via decentralization}
Besides functionality, trust is another powerful leverage to optimize the privacy-energy trade-off.
The PET literature usually considers two edge cases: (1) a party trusted by all users requiring no PET and (2) a zero-trust world requiring (potentially) expensive PETs.

However, decentralized social media such as Mastodon introduced a new kind of trust assumption: decentralized and personalized trust.
In these social media, each user can pick a specific server (that they trust to process their personal data), and then they can interact with any user (even from other servers) thanks to the protocol ActivityPub.
Such decentralization is comparable to the decentralization of the email protocol.

We can formulate the Mastodon threat model as follows: each user trusts one specific server (among all possible servers), and they do not trust anyone else.
Such decentralized trust avoids expensive cryptographic operations because each server can perform plaintext operations on the personal data of its users.
For instance, on Peertube (a ``decentralized Youtube''), each server can offer ML-based video recommendations to its users without relying on encrypted ML.

Decentralization ``\textit{à la} Mastodon'' is then a promising direction to enhance privacy while avoiding significant energy consumption overheads.

Mansoux and Roscam \cite{mansoux_seven_2020} presented this decentralized trust as ``social approach of privacy'', opposed to the technical approach adopted by cryptographic PETs.
While this ``social'' approach provides weaker privacy guarantees, Lee and Wang \cite{lee_uses_2023} showed that privacy was a key factor driving the adoption of Mastodon.
Thus, the weaker guarantees are not necessarily an issue for user acceptability (even among privacy-aware users).

However, one may argue that decentralized protocols require infrastructure redundancy as it requires many servers.
For example, Mastodon is a decentralized social media composed of 8K servers (in September 2025).
But, it does not imply that Mastodon requires 10K times more servers than a centralized social media.
Indeed, centralized social media commonly rely on a global network of servers.
Thus, both decentralized and centralized social media rely on a network of servers; the only difference is in the server ownership.
Thus, the subtle question of infrastructure and communication overheads in decentralized protocols must be careful studied because even large-scale centralized services rely on distributed systems creating redundancies and overheads.
This calls for dedicated future work (e.g., comparing Matrix and Signal protocols in secure messaging).

\section{Conclusion}
Our work studied the energy consumption of three cryptographic PETs with different levels of maturity.
Our results highlight PET-induced energy consumption increases ranging from $2\times$ (for TLS) to $100{,}000\times$ (for FHE).
This wide range also suggests a relationship between maturity and efficiency: the most mature technology is the most energy-efficient.

However, this correlation should not hide the fact that FHE remains several orders of magnitude more expensive than its plaintext equivalent.
While performance optimization is a central goal in cryptography research, this raises a key question: what is the optimal performance?
Modern FHE schemes rely on inherently expensive primitives, unlike TLS, which makes reaching comparable efficiency difficult, if not impossible.
This observation should not discourage research on FHE, but it illustrates a potential broader debate for the cryptography community about the trade-off between privacy and environmental footprint.



\begin{credits}
    \subsubsection{\ackname}
    This work was supported by the Netherlands Organization for Scientific Research (De Nederlandse Organisatie voor Wetenschappelijk Onderzoek) under NWO:SHARE project [CS.011]. We thank Ségolène Blanchard for proofreading.
    A Large Language Model was \textbf{carefully} used to polish some paragraphs.
\end{credits}

%

\bibliographystyle{splncs04}
\bibliography{ref}

\end{document}